\documentclass[RNAAS,natbib]{aastex63}

\begin{document}

\title{A Historical Perspective on the Diversity of Explanations for New Classes of Transient and Variable Stars}

\correspondingauthor{Thomas J. Maccarone}
\email{thomas.maccarone@ttu.edu}
\affiliation{Department of Physics and Astronomy, Texas Tech University, Lubbock TX 79409-1051}

\begin{abstract}
As new classes of transients and variable stars are discovered, and theoretical models are established to work or not to work for a few members of the class, it is often the case that some researchers will make arguments on the basis of Occam's razor that all members of the class must be produced by whichever mechanism first successfully explained one of the objects.  It is also frequent that this assumption will be more more implicitly.  Retrospective analysis shows rather clearly that this argument fails a large fraction of the time, and in many cases, this search for false consistency has led to more fundamental astrophysical errors, a few of which are quite prominent in the history of astronomy.  A corollary of this is that on numerous occasions, theoretical models to explain transients have turned out to be models that describe real (but often not yet discovered) phenomenona other than the ones to which they have first been applied, albeit with minor errors that caused the model to appear to fit to a known phenomenon it did not describe.  A set of examples of such events is presented here (some of which will be quite familiar to most astronomers), along with a discussion of why this phenomenon occurs, how it may be manifesting itself at the present time. Some discussion will also be made of why and when survey designs have led to immediate separation of various transient mechanisms, generally by being overpowered in some way relative to what is needed to {\it detect} a new class of objects.
\end{abstract}



\section{Introduction}

A repeated process in the history of astronomy is that the opening of new ``discovery space'' will lead to the discovery of new astrophysical phenomena (see \citealt{Harwit_book} for a rather comprehensive treatment of how this process has manifested itself with the introduction of new observational capabilities).  As this is done, often, classes will appear to emerge among the objects discovered.  Searches for new classes of transient events in the optical bandpass have come more into fashion in recent years than they had been for several decades prior.  In astronomy, as new classes of events and objects are discovered, they are ordinarily ``barely" discovered -- the first set of data that helps establish the new class is sufficient to establish clearly that {\it something} new has been found, but not to provide good diagnostics of the mechanism for productive the events or objects.  In this paper, I will discuss several historical examples of this process playing itself out, and show that in many occasions, these new phenomena represent hetereogeneous classes of objects.  Conversely, it is also often the case that models developed for explaining one set of phenomena have minor flaws that lead to them being correct descriptions of some other phenomenon later discovered, and I will present some examples of these, as well.  Finally, I will also discuss examples of cases where survey design helped identify the phenomenological differences that break the otherwise similar phenomena into two classes more easily.

\section{The multiplicity of nova mechanisms}
\subsection{The first samples}
The era of astrophotography began in the 1840s, with John Draper's successful daguerrotype of the Moon.  In the late 1800s and early 1900s, largely but not entirely making use of plates, systematic studies began to be made of optical transients.  It is instructive to look at an early table of what were all then called novae, containing 28 such objects mostly discovered between 1885 and 1912, but stretching back into the late 1500s \citep{1912AnHar..56..165F}.   This catalog neglects some earlier nova and supernova discoveries that have been recorded primarily by Chinese astronomers, for which the precision of the data in the historical record was deemed insufficient.  This list and a few addenda represented the state of the art for thought about the nature of optical transients for quite some time.  Within this rather small catalog exist supernovae of both Type Ia and Type II; classical and recurrent novae; a Mira~Ceti variable; and a luminous blue variable.  We present this catalog, along with current classifications for its members, in Table 1. 

Furthermore, in the year after this catalog was produced, WZ~Sge was discovered.  This is a very nearby dwarf nova, with a peak {\it apparent} magnitude similar to those of the classical novae known in the early 1900s.   The original discovery paper reported that while no spectrum was obtained during the outburst, ``Its sudden appearance, however, followed by a fading of brightness that was at first rapid and then more gradual, is in conformity with the typical light curve for novae" \citep{1919HarCi.219....1L}. Only five decades later, when its distance could be estimated based on its M-dwarf companion star, was it appreciated that its outburst was a clear clue  that this was something different than that of the classical novae\citep{1962ApJ...135..408K,1964ApJ...140..921K}. Other dwarf novae {\it had} been recognized as a different class of objects already, as U~Gem showed rapid recurrence after its original discovery \footnote{The object's discovery was first reported by J.R. Hind in a letter to {\it The Times} of London -- see \cite{1986MNSSA..45..117W}.}  \citep{1857MNRAS..17..200P}.  Thus at the time of the publication of \cite{1919HarCi.219....1L}, there was a class of about 30 events which were considered to be a homogeneous set of objects by most of the leading researchers of the era, but which we know understand to have {\it six} separate mechanisms -- thermonuclear detonation of the entirety of a white dwarf; core collapse driven explosion of a massive star; runaway nuclear fusion on the surface of a white dwarf; pulsation of a red giant; mass ejection by a blue supergiant; and a disk instability (perhaps coupled with enhanced mass transfer of the donor star) in an accretion disk around a white dwarf.

These objects showed a broad range of similarities, within the limits of the data of the era.  Where spectra could be obtained, with the exception of P~Cygni (the luminous blue variable), they appeared to be spectra of gaseous nebulae \citep{1912AnHar..56..165F}, and even P~Cygni still showed the emission lines for which it is so famous.  The characteristic decay times of the events were rather similar.  It is thus unsurprising, especially at the dawn of systematic academic study of transient sources, that people would have put all these objects into a single class.  Furthermore, it is probably desirable not to break objects up into an excessive number of subclasses until there is ample evidence to indicate that they are, indeed, caused by different phenomena.

\begin{table*}[]
    \centering
    \begin{tabular}{l|r|c|l|l}
    \hline
      Object name  &Disc yr. & $m_V$ &Class& Comment \\
      \hline
        B Cas & 1572& $-5$& Supernova& Tycho's\\
        S And & 1885& 7& Supernova& in M31\\
        V Per & 1887 & 9.2& classical nova&\\
        N Per 1901 & 1901& 0.0 & Classical nova & GK~Per\\
        T~Aur& 1891&4.5&classical nova & \\
        N Gem 1903& 1903 & 5.1& classical nova &\\
        N Vel 1905& 1905& 9.7&  classical nova &\\
        RS Carinae & 1895& 8 & classical nova\\
        Z Cen & 1895 & 8 & supernova& in NGC~5252\\
        Nova Cir& 1906& 9.5&classical nova& typo on year in \cite{1912AnHar..56..165F}\\
        R Norma&1893&7&Mira Ceti variable&\\
        T Cor B& 1866& 2& recurrent nova& red giant donor\\
        T Sco& 1860&7&classical nova& in globular cluster M80\\
        Nova Ara&1910&6&classical nova& \\
        N Oph 2&1848&5.5& classical nova&\\
        N Oph 1&1604&$-4$&supernova& Kepler's\\
        RS Oph&1898&7.7 &recurrent nova& red giant donor\\
        N Sco 2&1906& 8.8& classical nova&\\
        N Sgr 2 & 1910& 7.5& classical nova\\
        N Sgr 4 &1901&10.4& classical nova\\
        N Sgr 3& 1899& 8.5& classical nova\\
        N Sgr 1& 1898& 4.7& classical nova\\
        N Aql 2& 1905& 9.1& classical nova\\
        N Aql 1&1899&7& classical nova\\
        11 Vul &1670& 3& classical nova &CK~Vul\\
        P Cyg& 1600& 3.5& luminous blue variable&\\
        Q Cyg&1876&3&classical nova&\\
        N Lac&1910&5&classical nova&\\
        \hline
    \end{tabular}
    \caption{The 28 ``novae'' from \cite{1912AnHar..56..165F}.  The columns are the object name, the year of the discovery of the source, the peak apparent magnitude of the object, the class of object, and any comments that might be relevant to the objects.}
    \label{tab:my_label}
\end{table*}

\subsection{Separation of the supernova from the novae}
This growth in the size of ``nova" samples took place at the dawn of extragalactic astronomy.  The peak luminosities of novae were, in fact a core issue in the Shapley-Curtis debate of 1918 -- both participants agreed that novae had been seen in the spiral nebulae, with Shapley arguing that the novae would have to be perversely bright for S~And (now known as SN 1885, in M31) to be a nova if the spiral nebulae were island universes. Curtis, on the other hand, presciently argued that S~And might be a member of a different class of objects (a few years before the first really serious work on the topic was done -- \citealt{1923PASP...35...95L}), and that Tycho's nova was anomalously bright as well, and might be due to that same alternative mechanism \citep{1995PASP..107.1133T}.

In relatively short order this question was resolved by Hubble's discovery of Cepheids in nearby galaxies.  The Cepheids clearly established the spiral nebulae to be at $\sim$ Mpc distance scales, and hence that there was a class of nova dramatically brighter than the classical novae.  It also became clear that the large majority of nova events in M31 has luminosities compatible with those of bulk of the Galactic novae, while S~And was about 11 magnitudes brighter than the rest, and for more distant galaxies, it also became clear that a few of their novae were excessively bright given their distances \citep{1923PASP...35...95L}.  The term supernova was coined, and gradually, it became clear that this was a broad class of rare, bright objects with durations and light curve shapes not dramatically different from those of classical novae with the quality of light curves available in the early 20th century \citep{1934PNAS...20..254B,1938ApJ....88..285B}.  Notably, the same phenomenon occurred with the Cepheids, in the sense that the original scale of the Universe proposed by \citet{1929PNAS...15..168H} was incorrect due to the assumption that all Cepheids were the same, an assumption not rectified for quite some time \citep{1944ApJ...100..137B}.  

\subsection{Multiple mechanisms for supernovae}
The separation of the supernovae from the classical novae was eventually done in a straightforward manner using extragalactic events.  About two decades after the first well-developed suggestion\citep{1923PASP...35...95L} that the novae and supernovae were separate classes of objects, \cite{1941PASP...53..224M} showed that there appeared to be two separate classes of {\it supernovae}, based on their spectral dichotomy, with the Type II supernovae showing strong hydrogen lines, and the Type I showing no really strong spectral features in early data.   The Type Ib supernovae, which are core collapse supernova from Wolf-Rayet stars (i.e. stars without hydrogen envelopes) were identifed as peculiar Type I supernovae long after the Type I/Type II dichotomy was established but near the infancy of the development of serious theoretical models for supernovae \citep{1964AnAp...27..319B}.  

When the first distinct ideas for the two mechanisms were proposed \citep{1960ApJ...132..565H}, a paper which got the basic ideas for both models correct, but ironically, given the authors' fame for understanding cosmic nucleosynthesis, suggested that the Type II supernovae mostly produce iron, while the Type I's mostly produce alpha elements.  Only a quarter century later was it widely understood that the phenomenological boundary between Type I and Type II supernovae, in the absence or presence, respectively, of hydrogen lines, was not perfectly correlated with the explosion mechanism, and that the peculiar Type I supernovae (which we now call Type Ib and Type Ic) are core collapse supernovae from stripped stars \citep{1985ApJ...294L..17W}.

\subsection{Multiple mechanisms for dwarf novae}
Over time, mostly in the 1940s and 1950s, it was also realized that a subset of accreting white dwarfs showed ``dwarf novae" that were about 100,000 times fainter ($M_V$ typically around +5 -- \citealt{1987MNRAS.227...23W} ) than the classical novae (absolute magnitudes of about $M_V=-8$ -- \citealt{1981PASP...93..165D}).  For the dwarf novae, in which there are relatively sudden and extreme changes in the mass transfer rates through accretion disks around white dwarfs, there is also a breadth of phenomenology, with some attempts to explain things via a single mechanism.  The dwarf novae may be produced either by some instability in the accretion disk due to changes in the ionization state of the dominant species, usually hydrogen\footnote{In the case of ultracompact binaries, helium, or in some ultracompact X-ray binaries, carbon and oxygen may be the dominant species.}  \citep{{1981A&A...104L..10M},{1984AcA....34..161S}}, or variations in the mass transfer rate into the accretion disk \citep{1970ApJ...162..621O}.

For quite some time it has been clear that the ionization instability mechanism is important.  The evidence for this is particularly clear from the low mass X-ray binaries (i.e. black holes or neutron stars accreting from low mass donor stars), in which a clear demarcation is seen in a plot of mean accretion rate versus orbital period, and the systems predicted to be persistent in the disk instability model are, in fact, persistent \citep{2001bhbg.conf..149L}.  

Still, this does not exclude the idea that some outburst phenomenology may be driven by mass transfer variations.  The new generation of optical variability surveys have started to find relatively clear evidence that some of the outbursts of accreting white dwarfs cannot be explained purely by disk instabilities.  In \cite{2020arXiv201210356R}, SDSS~J113732+405458, a double white dwarf binary with an orbital period of about 60 minutes, showed a year-long outburst.  Not only is this timescale is longer than the viscous timescale for mass to flow through such a short period system's accretion disk, but also, the system became {\it redder} during the outburst, and maintained colors too red for the helium in the disk to be ionized, and the outburst was subluminous relative to standard dwarf novae, even for its short orbital period.  

\section{Gamma-ray bursts}

The cosmic gamma-ray bursts represent another example of a class of object first believed to be homogeneous, but later broken into at least three separate classes with fundamentally different mechanisms (magnetar outbursts, neutron star mergers, and fireballs produced during core collapse supernovae).  Interestingly, in her historical article about the first ``Great Debate", the Shapley-Curtis debate, \cite{1995PASP..107.1133T} noted a point of analogy to the then contemporary debate (celebrated on the Diamond Jubilee of the original) on the origin of gamma-ray bursts.  She pointed out that in 1995, there were suggestions being made that the gamma-ray bursts might be made up of two separate populations, as had been the case for the novae in Curtis' arguments in the original ``Great Debate.''\footnote{This was already long after the soft gamma repeaters had been identified as likely coming from a separate mechanism, due to their (as the name suggests) softer emission and repeated bursts, as well as the fact that the known objects are all either in the Galactic Plane or the Magellanic Clouds, indicating that they are produced by relatively nearby phenomena associated with massive stars.}

Still, for the broader class of gamma-ray bursts, the ones at cosmological distances, substantial numbers of members  of both classes had been observed before it was widely recognized that they were from two separate populations \citep{1993ApJ...413L.101K}.  The gamma-ray bursts were first reported in 1973, after having been detected by the Vela satellites from 1969-1972 \citep{1973ApJ...182L..85K}.  It was understood relatively quickly that the bursts occur with quite a large range on characteristic timescales, and suggestions existed that the distribution of timescales might be bimodal \citep{1984Natur.308..434N}, and it was well established that the bursts were isotropically and uniformly distributed \citep{1989ApJ...342..521H}.  Furthermore, claims of cyclotron lines and annihilation lines in the spectra of GRBs argued for highly magnetized neutron stars at modest redshift as a mechanism \citep{1982Ap&SS..82..261M,1988Natur.335..234M}.  

The isotropy led to a preference for extragalactic models, but the lines led to a preference for Galactic models, as the locations of the annihilation lines would be shifted in the event that the bursts were at cosmological distances.  Over time, skepticism grew about the observations themselves, in part because the detections of lines were not found to be repeatable with BATSE \citep{{1994ApJ...433L..77P}}.  For quite some time, though, the prevailing view was that the GRBs were Galactic sources, and this was partly driven by the putative spectral lines, along with concerns that, at cosmological distances, rampant pair production would prevent the spectra from appearing as they are (a problem solved by the strong beaming in these systems).  This combination of preference for Galactic neutron star models and the need to reproduce an isotropic sky distribution led to a variety of ``super-halo" models that managed to reproduce the isotropy via placing the GRBs in a large halo around the Milky Way, invoking neutron star kicks to do it (e.g. \citealt{1995PASP..107.1152L}).  There were, on the other hand, some researchers who pushed for the cosmological nature of the gamma-ray bursts, but again, in and effort to unify what should not have been unified, it was suggested that the soft gamma-ray repeaters might be lensed versions of the same phenomenon as the non-repeating GRBs \citep{1986ApJ...308L..43P}.

\section{General properties of new transient searches}
In most cases where a new window has been opened on the transient Universe, whether due to the use of a new waveband of electromagnetic radiation, new depth of sensitivity, or new cadence of observations, {\it some} new class of events has been discovered, with the exceptions being in TeV gamma-ray astronomy and sub-100 MHz radio astronomy.  Most of the time, when new capabilities are opened up, the increase in sensitivity is such that new phenomena can be discovered, but not necessarily well-characterized.  In transient searches, the cadence and duration of the survey, along with the sensitivity, will necessarily make new classes of objects look rather similar to one another unless there has been a dramatic expansion of the capabilities in detecting transients in {\it multiple} dimensions (from among, for example, depth, solid angle, cadence, duration of the survey, set of wavebands); it is usually the case that only one dimension of discovery space is dramatically improved, and hence new surveys typically will discover new classes of objects that look relatively similar to one another; events which are very different from these will either be ``too easy'' to detect (and hence will already have been discovered in past surveys) or ``too hard'' to detect (and hence not appear at all, or at least appear so infrequently as not to be recognized as a class).  

Now, we can ask what the consequences are of the facts both that (1) these new surveys are typically successful and (2) the characteristics of the transient events will necessarily be similar in most surveys.  It would be rather unlikely for a new region of discovery space to contain exactly a single class of event most of the time, unless physics prohibits any other mechanism from working in that corner of parameter space.  Where there is a substantial new class of objects, then, it is {\it more likely} to be hetereogeneous than to be homogeneous, especially if there exist several viable theoretical models for producing the sources.  If homogeneous new classes were the norm, it would follow that a lot of new projects would fail to discover anything unexpected.

\subsection{Occam's Blender and the hetereogeneity of explanations}
This conclusion can be troubling to people brought up with the idea that Occam's razor suggests that invoking two or more models to explain what appears to be a single phenomenon is fundamentally incorrect.  Still, the key qualification is that often, with relatively sparse data sets at the advent of the discovery of a new broad class of objects or events, there is often not sufficient data richness to distinguish among subclasses, and there is even less frequently sufficient data quality to identify the feature that can, in hindsight, be used to separate the objects.  In this situation, which frequently arises in astronomy, the most natural explanation for a new class of objects is that it is inhomogeneous; we propose adopting the term Occam's Blender\footnote{This term was coined by Prof. Dennis Ugolini at Trinity University after the author asked some friends and colleagues for suggestions of how to describe the situation.} for this concept.

A related phenomenon is that there are often ideas developed for, or incorrectly applied to, one astronomical phenomenon that re-appear to explain another phenomenon at a later time.  A classic example of this is the idea of supernova shock breakouts \citep{{1968CaJPS..46..476C,1974ApJ...187..333C}}.  This idea was proposed as a purely theoretical prediction, invoked as a gamma-ray burst model \citep{1974ApJ...187..333C}, and then rejected in the era of GRB afterglows.  It resurfaced when the actual shock breakouts, in the X-ray band, were discovered from supernovae \citep{2008Natur.453..469S}.

Similarly,  \cite{1976Natur.263..101W} proposed runway thermonuclear burning on the surface of a neutron star as a mechanism for the gamma-ray bursts.  The phenomenon is now known to occur in Nature, but to produce Type~I {\it X-ray} bursts, as the envelopes in which this material is burning are optically thick and convert the gamma-rays produced by the nuclear interactions into larger numbers of X-rays on the way out.  The Type~I X-ray bursts had been discovered at the time of \cite{1976Natur.263..101W}, in a paper by \cite{1972ApJ...171L..87B}, but the original interpretation was that the burst was a precursor to the outburst of Cen~X-4; in hindsight, it is now clear that the outburst started a few days before the Vela 5B satellite could detect it, and during the early phase of the outburst, enough matter was accreted to trigger a thermonuclear burst. Discoveries of additional Type I bursts \citep{1976ApJ...205L.127G,1976ApJ...206L.135B} came a few months before the paper by \cite{1976Natur.263..101W} was published.  Interestingly, if it had been realized immediately that the event discovered by \cite{1972ApJ...171L..87B} was from the same class as the events discovered by \cite{1976ApJ...205L.127G}, it may have led to a more immediate adoption of the nuclear fusion model; the alternative of model of scattering in a cloud of hot gas around an intermediate mass black hole \cite{1976ApJ...205L.131G} relied on the idea that these events had seemed to occur only in globular clusters.  Regardless of the early (and understandable) confusion, was still the case that by the late 1970s/early 1980s, numerous papers were written discussing the nuclear burning model in great detail (e.g. \citealt{1978ApJ...225L.123J,1980ApJ...241..358T}), while very little further discussion was given to the intermediate mass black hole idea. 

 Retrospective examination of the gamma-ray burst models also indicates that one shouldn't get carried away and assume that all theoretical work is correct just because a few of the GRB models turned out later to be relevant for other phenomena (or similarly, because there is now evidence that both of the mechanisms first proposed for dwarf nova outbursts do, sometimes, work).  It is also clear from the old literature that not every seemingly viable model for a phenomenon actually happens with substantial frequency -- e.g. there is, with the benefit of hindsight no reason to believe that gamma-ray bursts represent interstellar nuclear warfare nor Oort Cloud anti-matter comets nor many of the other less exotic, but still incorrect GRB models (see \citealt {1994ComAp..17..189N} for a rather extensive tabulation of models some of which are more likely than other eventually to be connected to rarer classes of astrophysical transient).

\section{The present and the future}
Over the past few decades, a wide variety of new classes of objects have been discovered (or have seen samples grow in size from a few oddball objects to a real class of objects).   Other debates started many decades ago have remained unresolved (e.g. whether the Type Ia supernovae are produced via a single or double degenerate channel, and whether the single degenerate channel can include garden variety cataclysmic variables or only symbiotic stars with persistent supersoft emission).  For the open questions, where there are multiple viable models that do not yet make any robustly testable distinct predictions, it is likely that more than one of them will turn out to be correct in many cases.  This has been seen with some of the more recent newly discovered classes of objects, and remains to be determined for others.

\subsection{New hetereogeneous classes}
Ultraluminous X-ray sources are objects not associated with the nuclei of giant galaxies and which have X-ray luminosities greater than $10^{39}$ erg/sec.  In recent years, a few of these have shown X-ray pulsations, indicative of a hyperaccreting, highly magnetized neutron star \citep{{2014Natur.514..202B}}.  Still, there are several such objects which have properties that are problematic for the idea that {\it all} ULXs are accreting high magnetic field neutron stars.  In particular, some of the ULXs are known to be associated with globular clusters, where high magnetic fields are very unlikely (e.g. \citealt{2007Natur.445..183M,2019MNRAS.485.1694D}), while another prominent object shows state transitions and variable radio emission, characteristic of a scaled-up stellar mass black hole \citep{2012Sci...337..554W}.  The ULXs are highly likely to represent a mixture of accreting high magnetic field neutron stars and accreting black holes of both stellar and intermediate mass.

At the other end of the luminosity range,the very faint X-ray transients \citep{2005ApJ...633..228M,2009A&A...495..547D} also represent a class of objects that was quickly established to be heterogeneous after its discovery.  These are X-ray transients that are bright enough that they are highly likely to be predominantly black hole or neutron star accretors, but faint enough that they do not trigger all-sky monitors unless they are very nearby.  They are likely to represent the bulk of X-ray binaries, given their relatively large numbers in the small patches of the sky in which careful searches have been made.  The early discovery that one of the objects was eclipsing \citep{2005ApJ...633..228M}, along with the long-established knowledge that eclipsing X-ray binaries are underluminous, because only a component scattered in the disk wind is seen, helped set the stage for the idea that this was a heteregeneous class of objects.  Now, as more of these objects have been identified, it is very clear that they are quite a mixed class, including some foreground accreting white dwarfs among them \citep{2020MNRAS.492.4344S}.

\subsection{New mysteries that are likely to be heteregeneous}
Some other unsolved mysteries include the mechanisms for producing fast radio bursts, and the channels by which the double black hole mergers seen in LIGO data are produced.  The historical evidence suggests that the use of the plural in the preceding sentence is appropriate -- these are probably both heterogeneous.  For the fast radio bursts, dozens of theoretical models have been proposed\footnote{https://frbtheorycat.org/index.php/Main\_Page}.  It would be surprising if only one of them is valid.  Furthermore, there are already FRBs which repeat frequently, and those that either do not repeat or repeat only infrequently.  All that is known, fundamentally, about the FRBs as a class (as opposed to the relatively small subset that has been localized) is that they are extragalactic and produce fast, short bursts of radio emission,and that their discovery has been enabled in part by new hardware, but in part by people undertaking the computational effort to search for single pulses over a wide range of dispersion measures in large data sets \citep{{2007Sci...318..777L}}.

For the gravitational wave sources, it is about as clear as anything astrophysical can be {\it what} they are, as the discovery data themselves give precise information on the final states of the systems; what is more uncertain is how they formed.  Indeed, in recent years, four major hypotheses have been developed -- standard binary stellar evolution (e.g. \citealt{2016ApJ...819..108B}); chemically homogeneous binary stellar evolution (e.g. \citealt{2016MNRAS.460.3545D}); dynamical formation in globular clusters(e.g. \citealt{2016PhRvD..93h4029R}); and formation in the accretion disks of active galactic nuclei (e.g. \citealt{2018ApJ...866...66M}).  At the present time, some aspects of some of the sources' data are challenging to some of the models.  For example, chemically homogeneous evolution is likely to produce large spins, and at least some of the detected sources appear to have small spins; this does not exclude the idea that the chemically homogeneous evolution channel may produce some substantial fraction of the events.  Fortunately, there is a growing acknowlegment of the idea that heterogeneity of explanations is likely even in papers that argue that it is possible for one of the models to explain everything (e.g. \citealt{{2021RNAAS...5...19R}}).

\subsection{Avoiding this problem in the future?}
Ideally, one wants to set up the discovery process in a manner that allows immediate separation of different classes of transients.  In some cases the discovery data lend themselves to a clean and correct interpretation almost immediately, while for other phenomena, the root cause has remained elusive for a much longer period of time -- see \cite{2006AIPC..836....3T} for a discussion of a large set of examples of the range of timescales from discovery to understanding in astronomy.  

In some of the most fortunate cases (like the discovery of radio pulsars) the manner in which the discovery data were collected led in a relatively straightforward manner to the interpretation \citep{1968Natur.217..709H}.  Still, it could have easily gone wrong; if we imagine a world in which Jocelyn Bell had not looked so closely at the strip charts reading her data, the pulsars might first have been recognized as a separate class of objects not via their rapid periodic variability, but via their radio spectra, which are much steeper than most, but not all other radio sources.  The availability of time resolution in the data well beyond what was require for the initial goals of the project -- detecting scintillation due to the interstellar medium -- paved the way for the discovery and nearly immediate interpretation of the pulsars.  The soft gamma-ray repeaters, as the name suggests, were quickly distinguished from other gamma-ray bursts because of their repetition, and, to some extent, their softer spectra.   Similarly, when a second class of {\it X-ray} bursts, imaginatively named Type II X-ray bursts, was discovered in the ``Rapid Burster," it was immediately recognized as something qualitatively different \citep{1976ApJ...207L..95L}, and this quickly led to the still-favored interpretation of magnetic gating of the accretion flow.  

These examples above are all cases where the discovery data set was ``overpowered" relative to what was needed to notice the phenomenon.  Pulsars would have been found to scintillate, and to have unusual radio spectra, even without the high time resolution of the strip chart data.  The fact that the data recording method had time resolution far in excess of what was needed for the project's aims, coupled with a particularly alert graduate student working on the project, led to the discovery of pulsars.    For high energy observations, there can be extremely rapid, extremely high amplitude phenomena coupled with nearly continuous coverage of the whole sky.  In these cases, durations of observations far exceed what is needed to make the discoveries.  In more recent years, the Palomar Transient Factory project, by aiming for wide and shallow observations, so that the imaging does not overshoot the capability for doing spectroscopic follow-up, has generally been very successful at identifying new classes of transients, even when their light curves are relatively similar (e.g. \citealt{2012ApJ...755..161K}).  {\it Swift} represents another project for which the set of instruments was well designed, such that new classes of transients could be immediately identified.

This is notably {\it not} the case for the Vera Rubin Observatory's Legacy Survey of Space and Time -- its detection limits will dramatically overshoot the resources available for spectroscopic follow-up at those fluxes, and its light curves, except in a few drilling fields, will be sampled with quite a low duty cycle, and in relatively few filters.  Some potential exists to improve the discovery space if the LSST data are supplemented with e.g. strong time domain capabilities at radio and X-ray bands.  Still it is {\it very} likely that the early stages of LSST will yield new classes of transients which are hard to follow-up, and where multiple distinct mechanisms for producing the transients have rather similar observational appearance.  

This should be borne in mind ahead of the survey's commencement in two ways.  First, researchers should be aware that the new classes are likely to be hereteogeneous.  Secondly,  complimentary facilities to characterize these transients based on their multiwavelength behavior (e.g. at radio and at high energies) should be supported.

\section{Acknowledgements}
I thank Chris Fryer, Nathalie Degenaar, Jean in't Zand, Rudy Wijnands, Aarran Shaw, Craig Heinke, Liliana Rivera Sandoval, Kristen Dage, \v{Z}eljka Bo\v{s}njak and Selma de Mink for enlightening discussions, some directly motivated by this paper, and some tangential to it, that helped foment the ideas in this paper and motivate its writing.  I also thank the referee for comments which improved the paper.

\end{document}